\documentclass[10pt, conference, compsocconf]{IEEEtran}

\usepackage{algorithm}
\usepackage{algpseudocode}
\usepackage{makeidx}  
\usepackage[dvips]{graphicx}
\usepackage{booktabs}
\usepackage{url}
\usepackage{amssymb}
\usepackage{amsmath}
\usepackage{color}
\usepackage{epstopdf}
\newtheorem{theorem}{Theorem}

\newtheorem{definition}{Definition}

\usepackage{calrsfs}
\usepackage{fourier} 
\DeclareMathAlphabet{\pazocal}{OMS}{zplm}{m}{n}
\newcommand{\Ia}{\mathcal{I}}
\newcommand{\Ja}{\mathcal{J}}

\newcommand{\BigO}[1]{\ensuremath{\mathcal{O}(#1)}}

\begin{document}
	
\title{Minimizing Total Busy Time with Application to Energy-efficient Scheduling of Virtual Machines in IaaS clouds}

\author{\IEEEauthorblockN{Nguyen Quang-Hung, Nam Thoai}\\
	\IEEEauthorblockA{Faculty of Computer Science and Engineering, \\
		HCMC University of Technology, VNU-HCM \\
		Ho Chi Minh City, Vietnam\\
		Email: \{hungnq2,nam\}@cse.hcmut.edu.vn}
}

\maketitle              


\begin{abstract}
Infrastructure-as-a-Service (IaaS) clouds have become more popular enabling users to run applications under virtual machines.
Energy efficiency for IaaS clouds is still challenge.
This paper investigates the energy-efficient scheduling problems of virtual machines (VMs) onto physical machines (PMs) in IaaS clouds along characteristics:
multiple resources, fixed intervals and non-preemption of virtual machines. The scheduling problems are NP-hard. 
Most of existing works on VM placement reduce the total energy consumption by using the minimum number of active physical machines.
There, however, are cases using the minimum number of physical machines results in longer the total busy time of the physical machines.
For the scheduling problems, minimizing the total energy consumption of all physical machines is equivalent to minimizing total busy time of all physical machines. 
In this paper, we propose an scheduling algorithm, denoted as EMinTRE-LFT, for minimizing the total energy consumption of physical machines in the scheduling problems.
Our extensive simulations using parallel workload models in Parallel Workload Archive show that the proposed algorithm has the least total energy consumption compared to the state-of-the-art algorithms.

\end{abstract}

\begin{keywords}
energy efficiency; energy-aware; power-aware; vm placement; IaaS; total busy time; fixed interval; fixed starting time; scheduling
\end{keywords}

\section{Introduction}
\label{sec:intro}
Infrastructure-as-a-Service (IaaS) cloud \cite{Zhang2010a} service provisions users with computing
resources in terms of virtual machines (VMs)
to run their applications \cite{Garg2009,Le2011,Beloglazov2012}.
These IaaS cloud systems are often built from virtualized data centers.
Power consumption in a large-scale data centers requires multiple megawatts \cite{Fan2007a,Le2011}. 
Le et al. \cite{Le2011} estimate the energy cost of a single data center is more than \$15M per year. 
As these data centers has more physical servers, they will consume more energy.
Therefore, advanced scheduling techniques for
reducing energy consumption of these cloud systems are highly
concerned for any cloud providers to reduce energy cost. 
Energy efficiency is an interesting research topic in cloud systems.
Energy-aware scheduling of VMs in IaaS cloud is still challenging \cite{Garg2009,Le2011,Quang-Hung2014,Tako2012}.

Many previous works \cite{Flammini2010, WTianMFFDE2013} proved that the scheduling problems with fixed interval times are NP-hard.
They \cite{Beloglazov2012,Panigrahy2011} present
techniques for 
consolidating virtual machines in cloud data centers by using 
bin-packing heuristics (such as First-Fit Decreasing \cite{Panigrahy2011}, and/or Best-Fit Decreasing \cite{Beloglazov2012}).
They attempt to minimize the number of running physical machines
and to turn off as many idle physical machines as possible. 
Consider a $d$-dimensional resource allocation
where each user requests a set of virtual machines (VMs). 
Each VM requires multiple resources (such as CPU, memory, and IO)
and a fixed quantity of each resource at a certain time interval.
Under this scenario, using a minimum of physical machines can result in increasing the total busy time of the active physical machines \cite{HungFDSE2014}\cite{WTianMFFDE2013}.
In a homogeneous environment where all physical servers are identical,
the power consumption of each physical machine is linear to its CPU utilization \cite{Beloglazov2012},
i.e.,
a schedule with longer working time will consume more energy
than another schedule with shorter working time.



This paper presents a proposed heuristic, denoted as EMinTRE-LFT, to allocate VMs that request multiple resources in the fixed interval time and non-preemption into physical machines to minimize total energy consumption of physical machines while meeting all resource requirements.
Using numerical simulations, we compare EMinTRE-LFT
with the state-of-the-art algorithms include Power-Aware Best-Fit Decreasing (PABFD) \cite{Beloglazov2012}, vector bin-packing norm-based greedy (VBP-Norm-L2) \cite{Panigrahy2011}, 
and Modified First-Fit-Decreasing-Earliest (Tian-MFFDE) \cite{WTianMFFDE2013}.
Using three parallel workload models  \cite{feitelson1996workloadmodel}, \cite{downey1998parallel} and \cite{lublin2003workload} in the Feitelson's Parallel Workloads Archive \cite{feitelsonpwa},
  the simulation results show that the proposed EMinTRE-LFT can reduce the total energy consumption of the physical servers by average of 23.7\% 
  compared with Tian-MFFDE \cite{WTianMFFDE2013}. 
  In addition, EMinTRE-LFT can reduce the total energy consumption of the physical servers by average of 51.5\% and respectively 51.2\% compared with PABFD \cite{Beloglazov2012} and VBP-Norm-L2 \cite{Panigrahy2011}.
Moreover, EMinTRE-LFT has also less total energy consumption than MinDFT-LDTF \cite{HungFDSE2014} in the simulation results. 

The rest of this paper is structured as follows. 
Section \ref{sec:related} discusses related works.
Section \ref{sec:problem} describes the energy-aware VM allocation problem with multiple requested resources, fixed starting and duration time. 
We also formulate the objective of scheduling, and present our theorems. 
 The proposed EMinTRE-LFT algorithm presents in Section \ref{sec:algoEMinTRE}.
 Section \ref{sec:experiment} discusses our performance evaluation using simulations.
 Section \ref{sec:concl} concludes this paper and introduces future works.
\section{Related Works}
\label{sec:related}

The interval scheduling problems have been studied for many years with objective to minimizing total busy time.
In 2007, Kovalyov et al. \cite{kovalyov2007fixed} has presented work to describe characteristics of a fixed interval scheduling problem 
in which each job has fixed starting time, fixed processing time, and is only processed in the fixed duration time on a available machine. 
The scheduling problem can be applied in other domains. 
Angelelli et al. \cite{Angelelli20113650} considered interval scheduling with a resource constraint in parallel identical machines. 
The authors proved the decision problem is NP-complete if number of constraint resources in each parallel machine is a fixed number greater than two.
Flammini et al. \cite{Flammini2010} studied using new approach of minimizing total busy time to optical networks application. 
Tian et al. \cite{WTianMFFDE2013} proposed a Modified First-Fit Decreasing Earliest algorithm, denoted as Tian-MFFDE, for placement of VMs energy efficiency. The Tian-MFFDE sorts list of VMs in queue order by longest their running times first) and places a VM (in the sorted list) to any first available physical machine that has enough VM's requested resources. 
Our VM placement problem differs from these interval scheduling problems \cite{kovalyov2007fixed}\cite{Angelelli20113650}\cite{WTianMFFDE2013}, 
where each VM requires for multiple resource (e.g. computing power, physical memory, network bandwidth, etc.) 
instead of all jobs in the interval scheduling problems are equally on demanded computing resource 
(i.e. each physical machine can process the maximum of $g$ jobs in concurrently). 

Energy-aware resource management in cloud virtualized data centers is critical. 
Many previous research \cite{Beloglazov2012,Knauth2012,Tako2012,Chen2014} proposed algorithms that consolidate VMs 
 onto a small set of physical machines (PMs) in virtualized datacenters 
 to minimize energy/power consumption of PMs.
A group in Microsoft Research \cite{Panigrahy2011} has studied first-fit decreasing (FFD) based heuristics for vector bin-packing to minimize number of physical servers in the VM allocation problem.
 Some other works also proposed meta-heuristic algorithms to minimize the number of physical machines.
Beloglazov's work 
 \cite{Beloglazov2012} has presented a modified best-fit decreasing heuristic in bin-packing problem, denoted as PABFD, 
 to place a new VM to a host.
PABFD sorts all VMs in a decreasing order of CPU utilization and tends to
allocate a VM to an active physical server that would take the minimum
increase of power consumption. 
Knauth et al. \cite{Knauth2012} proposed the OptSched scheduling algorithm to 
 reduce cumulative machine up-time (CMU) by 60.1\% and 16.7\% in comparison to a round-robin and First-fit.
 The OptSched uses an minimum of active servers to process a given workload.
 In a heterogeneous physical machines, the OptSched maps a VM to a first available and the most powerful machine that has enough VM's requested resources. Otherwise, the VM is allocated to a new unused machine. 
  In the VM allocation problem, however, minimizing the number of used physical machines is not equal to minimizing total of total energy consumption of all physical machines.
Previous works do not consider multiple resources, fixed starting time and non-preemptive duration time of these VMs.
Therefore,
it is unsuitable for the power-aware VM allocation
considered in this paper,
i.g. these previous solutions can not result in a minimized total energy consumption
for VM placement problem with certain interval time while still
fulfilling the quality-of-service.


Chen et al \cite{Chen2014} observed there exists VM resource utilization patterns.
 The authors presented an VM allocation algorithm to consolidate complementary VMs with spatial and temporal-awareness in physical machines.
 They introduce resource efficiency and use norm-based greedy algorithm, which is similar to in \cite{Panigrahy2011}, to measure 
 distance of each used resource's utilization and maximum capacity of the resource in a host. 
 Their VM allocation algorithm selects a host that minimizes the value of this distance metric to allocate a new VM.
 Our proposed EMinTRE-LFT uses a different metric that unifies both increasing time and the $L2$-norm of diagonal vector that is presenting available resources. 
 In our proposed TRE metric, the increasing time is the difference between two total busy time of a PM after and before allocating a VM. 

Our proposed EMinTRE-LFT algorithm that
differs from these previous works. Our EMinTRE-LFT algorithm use the VM's fixed
starting time and duration to minimize the total busy
time on physical machines, and consequently minimize the total
energy consumption in all physical servers. 
To the best of our knowledge, no existing works that surveyed in \cite{BelBuLZA2010Taxonomy,orgerie2014survey,Hameed2014,IvonaSurvey2014} have thoroughly considered these aspects in addressing the problem of VM placement.


\section{Problem Description}
\label{sec:problem}
\subsection{Notations}
We use the following notations in this paper:

$ vm_{i}$: The $i^{th}$ virtual machine to be scheduled.

$ M_{j}$: The $j^{th}$ physical machine.

$ S $: A feasible schedule. 

$ P_{j}^{min}$: The minimum power consumed when $M_{j}$ is 0\% CPU utilization.

$ P_{j}^{max} $: The maximum power consumed when $M_{j}$ is 100\% CPU utilization.

$ P_{j}(t) $: Power consumption of $M_{j}$ at a time point $t$.


$ ts_{i}$: Fixed starting time of $vm_{i}$.

$ d_{i}$: Duration time of $vm_{i}$.

$ T $: The maximum schedule length, which is the time that the last virtual machine will be finished.


$\Ja_j$: Set of virtual machines that are allocated to $M_j$ in the whole schedule.

$T_{j}^{busy}$ : The total busy time (ON time) of $M_{j}$.

 $ e_{i} $: Energy consumption for running $ vm_{i}$ in the physical machine that $ vm_{i}$ is allocated.
 
 $g$: The maximum number of virtual machines that can be assigned to any physical machine.\\

\subsection{Power consumption model}

Notations: \\ 
 - $ U_{j}(t)$ is the CPU utilization of $M_{j}$ at time $t$.
 - $ PE_{j}$ is the total number cores of $M_{j}$.\\
 - $ mips_{i,c} $	is the allocated MIPS of the $c^{th}$ processing element to the $vm_{i}$ by $M_{j}$.\\
 - $ MIPS_{j,c} $ is the maximum computing power (in MIPS) of the $c^{th}$ core on $M_{j}$.\\
 
In this paper, we use the following energy consumption model proposed in \cite{Fan2007a}\cite{Beloglazov2012} for a physical machine.
Let call $\alpha = P_{j}^{min} / P_{j}^{max}$ is fraction of the minimum power consumed when $M_j$ is idle (0\% CPU utilization) and the maximum power consumed when the physical machine is fully utilized (100\% CPU utilization).  The power consumption of $M_j$, denoted as $ P_{j}(.)$ with ($j = 1, 2,..., m$), is formulated as follow: 
\begin{equation}
\label{eq:power}
P_{j}(t) = (\alpha +(1 - \alpha).U_{j}(t)).P_{j}^{max}
 \end{equation}
  
We assume that all cores in CPU are homogeneous, i.e. $\forall c=1,2,...,PE_j: MIPS_{j,c}=MIPS_{j,1}$. The CPU utilization $ U_{j}(t)$ is formulated as follow:
 \begin{equation}
 \label{eq:cpuutilization}
 U_{j}(t) = (\dfrac{1}{PE_{j} \times MIPS_{j,1}}) \sum_{c=1}^{PE_{j}} \sum_{vm_i  \in  \Ja_j} mips_{i,c} 
 \end{equation}

The energy consumption of the $M_{j}$ in the time period [$t_1, t_2$] denoted as $\Delta E_{j}$ with CPU utilization $U_j$ is formulated as follow:
\begin{equation}
\label{eq:energy}
\Delta E_{j} = P_{j}( U_{j}).(t_2 - t_1)
	= ( \alpha . P_{j}^{max} + (1 - \alpha) . P_{j}^{max} . U_j ) . \Delta T
\end{equation}

{\noindent}where: \\
 $\Delta T_j$ : The busy time of $M_j$ that is defined as: $\Delta T_j = (t_2 - t_1)$.

Assume that a virtual machine $vm_i$ changes the CPU utilization is $\Delta u_j$ for during [$t_1, t_2$] and the $vm_i$ uses full utilization of its requested resources in the worst case on $M_j$. The energy consumption by the $vm_i$, denoted as $e_i$, is formulated as:
\begin{equation}
e_i = (1 - \alpha) . P_{j}^{max} . \Delta u_j  . (t_2 - t_1)
\end{equation}

Let $T_{j}^{busy}$ be the total busy time of $M_j$, let $e_{i}$ be energy consumed by $vm_i$, and  let $vm_i \in M_j$ be set of virtual machines $vm_i$ ($i = 1,2,..., n$) that are allocated to $M_j$ in the whole schedule. Let $E_j$ be the total energy consumed by $M_j$ and $E_j$ is the sum of energy consumption $\Delta E_j$ during the total busy time $T_{j}^{busy}$ that is formulated as: 
\begin{equation}
E_{j} = ( \alpha . P_{j}^{max} + (1 - \alpha) . P_{j}^{max} . U_j ) . T_{j}^{busy}
\end{equation}

where $\alpha . P_{j}^{max} . T_{j}^{busy} $ is called the base (ON) energy consumption for $M_j$ during the total busy time, i.e., $E_j^{base} = \alpha . P_{j}^{max} . T_{j}^{busy}$, and $((1 - \alpha) . P_{j}^{max} . U_j . T_{j}^{busy})$ is the increasing energy consumed by some VMs scheduled to $M_j$.

\begin{equation}
\label{eq:energytbt}
E_{j} =  \alpha . P_{j}^{max} \times T_{j}^{busy}  +  \sum_{vm_i \in M_j} e_{i}
\end{equation}

\subsection{Problem formulation}

Consider the following scheduling problem. 
We are given a set of $n$ virtual machines $ \Ja = \{ vm_1, \ldots ,vm_n \}$ to be scheduled on a
set of $m$ identical physical servers $\mathcal{M} = \{ M_1, \ldots , M_m \}$,
each server can host a maximum number of $g$ virtual machines. 
Each VM needs $d$-dimensional demand resources in a fixed interval with non-migration. 
Each $vm_{i}$ is started at a fixed starting time ($ts_{i}$) 
and is non-preemptive during its duration time ($d_{i}$). 
Types of resource considered in the problem include 
 computing power (i.e., the total Million Instruction Per Seconds (MIPS) of all cores in a physical machine),
 physical memory (i.e., the total MBytes of RAM in a physical machine), network bandwidth (i.e., the total Kb/s of network bandwidth in a physical machine), and storage (i.e., the total free GBytes of file system in a physical machine), etc.

The objective is to find out a feasible schedule $S$ that minimizes
the total energy consumption
in the equation (\ref{eq:minimize}) 
 with $\forall i \in \{1,2,...,n\}$, $\forall j \in \{1,2,...,m\}$, $\forall t \in [0,T]$ as following:

\begin{equation}
\label{eq:minimize-hete}
\textbf{Min} ( \sum_{j=1}^{m} ( \alpha \times P_{j}^{max} \times T_{j}^{busy} )  +  \sum_{i=1}^{n} e_{i} )
\end{equation}  

{\noindent}where: 

- $\alpha = P_j^{min} / P_j^{max} $ is the fraction of idle power and maximum power consumption by physical machine $M_j$. 
- $T_{j}^{busy}$ is the total busy time of $M_j$. 
%


In homogeneous physical machines (PMs), all PMs have the same idle power and maximum power consumption. Therefore $\alpha$ is the same for all PMs. We rewrite the objective scheduling as following: 
\begin{equation}
\label{eq:minimize}
\textbf{Min} ( \alpha \times P^{max} \times \sum_{j=1}^{m} T_{j}^{busy}  +  \sum_{i=1}^{n} e_{i} )
\end{equation} 

The scheduling problem has the following hard constraints that are described in our previous work \cite{HungFDSE2014} as following:
\begin{itemize}
	\item Constraint 1: Each VM is only processed by a physical server at any time with non-migration and non-preemption.
	
	\item Constraint 2: 
	Each VM does not request any resource larger than the maximum total capacity resource of any physical server.
	
	\item Constraint 3: 
	The sum of total demand resources of these allocated VMs is less than or equal to the total capacity of the resources of $M_j$.
\end{itemize}
\subsection{Preliminaries}

\begin{definition}[Length of intervals.] 
Given a time interval $I=[s, f]$, the length of I is $len(I) = f - s$.
 Extensively, to a set $\Ia$ of intervals, length of $\Ia$ is $len(\Ia) = \sum_{I \in \Ia} len(I)$.
\end{definition}


\begin{definition}[Span of intervals.]
 For a set $\Ia$ of intervals, we define the span of $\Ia$ as $span(\Ia) = len(\bigcup \Ia)$. 
\end{definition}




\begin{definition}[Optimal schedule]
An optimal schedule is the schedule that minimizes the total busy time of physical machines. For any instance $\Ja$ and parameter $g \geqslant 1$, $OPT(\Ja, g)$ denotes the cost of an optimal schedule.
\end{definition}


In this paper, we denote $\Ja$ is set of time intervals that derived from given set of all requested VMs. In general, we use instance $\Ja$ is alternative meaning to a given set of all requested VMs in context of this paper.

\textbf{Observations: Cost, capacity, span bounds. }
\label{def:capacitybounds}
 For any instance $\Ja$, which is set of time intervals derived from given set of all requested VMs, and capacity parameter $g \geqslant 1$, which is the maximum number of VMs that 
 can be allocated on any physical machine,
 the following bounds are held:\\ 
  $\bullet$ The optimal cost bound: $OPT(\Ja, g) \leq len(\Ja)$. \\ 
 $\bullet$ The capacity bound: $OPT(\Ja, g) \geqslant \dfrac{len(\Ja)}{g}$. \\ 
 $\bullet$ The span bound: $OPT(\Ja, g) \geqslant span(\Ja)$.\\

For any feasible schedule $s$ on a given set of virtual machines, the total busy time of all physical machines that are used in the schedule $s$ is bounded by the maximum total length of all time intervals in a given instance $\Ja$.
Therefore, the optimal cost bound holds because $OPT(\Ja, g) {=} len(\Ja)$ iff all intervals are non-overlapping, 
i.e., $\forall I_{1}, I_{2} \in \Ja$ then $I_{1} \cap  I_{2} = \emptyset $.

Intuitively, the capacity bound holds because $OPT(\Ja, g) {=} \dfrac{len(\Ja)}{g}$
iff, for each physical server, exactly $g$ VMs are neatly scheduled in that physical server.
The span bound holds because at any time $t \in \bigcup \Ja$ at least one machine is working.

\subsection{Theorems}
\label{sec:theorems}

In the following theorems, all physical machines are homogeneous. 
Let $P^{min}$ and $P^{max}$ are the minimum/idle power and maximum power consumption of a physical machine respectively. We have $\alpha = P^{min} / P^{max}$.
\begin{theorem}
\label{sec:theorem02}
 Minimizing total energy consumption in (\ref{eq:minimize}) is equivalent to
 minimizing the sum of total busy time of all physical machines ($\sum_{j=1}^{m} T_{j}^{busy}$).
 
\begin{equation}
\label{eq:minimize2}
\textbf{Min} \   ( \alpha \times P^{max} \times \sum_{j=1}^{m} T_{j}^{busy}  +  \sum_{i=1}^{n} e_{i} ) ~ \sim ~ \textbf{Min} \   ( \sum_{j=1}^{m} T_{j}^{busy})
\end{equation}

\end{theorem}

\begin{proof}
	A proof for this theorem see detail in \cite{HungFDSE2014}.
\end{proof}

Based on the above theorem, we propose our energy-aware algorithms denoted as EMinTRE-LFT which is presented in the next section.

\begin{definition}
\label{def:costtotalbusytime}
For any schedule we denote by $\Ja_j$ the set of virtual machines allocated to the physical machine $M_j$ by the schedule. 
 Let $T_j$ denote the total busy time of $M_j$ is the span of $\Ja_j$, i.e., $T_j = span(\Ja_j)$.
 \end{definition}
 
\begin{definition}
	\label{def:costalg}
For any instance $\Ja$, the total busy time of the entire schedule of $\Ja$ computed by the algorithm $H$, denoted as $cost^H(\Ja)$, is defined as
 $cost^H(\Ja) = \int^{span(\Ja)}_0 N^{H} (t) dt$,
 where as $N^{H} (t) $ is the number of physical machines used at the time $t$ by the algorithm $H$.
\end{definition}

\begin{definition}
\label{def:minenergy}
For any instance $\Ja$ and parameter $g \geqslant 1$, $E^{OPT}(\Ja, g)$, which is denoted as the minimized total energy consumption
 of all physical machines in an optimal schedule for the $\Ja$, is formulated as:   
 $E^{OPT}(\Ja, g) = \alpha \times P^{max} \cdot OPT(\Ja, g)  + \sum_{i=1}^{n} e_i $. \\
\end{definition}

\begin{theorem}
	For any instance $\Ja$, the lower and upper of the total energy consumption in an optimal schedule are bounded by: 	
	$P_{min} \cdot \frac{len(\Ja)}{g} \leq E^{OPT}(\Ja, g) \leq P^{max} \cdot len(\Ja) $. \\
\end{theorem}

\begin{proof}
	For any instance $\Ja$, let $OPT(\Ja, g)$ be the total busy time of the optimal schedule for the $\Ja$, 
	and let $E^* $ be the total energy consumption for the optimal schedule for the $\Ja$. 
	
	The total energy consumption of an optimal schedule needs to account for all physical machines running during $OPT(\Ja, g)$. We have:
	$E^* =  P_{min} \cdot OPT(\Ja, g)  + \sum_{i=1}^{n} e_i $. 
	
	From Definition \ref{def:minenergy}, we have $E^{OPT}(\Ja, g) = E^*$. 
	
	Apply the capacity bound in Theorem \ref{def:capacitybounds}, we have $OPT(\Ja, g) \geq \frac{ len(\Ja)}{g}$. Thus, 
	$E^* \geq  P_{min} \cdot \frac{len(\Ja)}{g}  + \sum_{i=1}^{n} e_i$.

	Recall that the energy consumption of each virtual machine is non-negative, thus $e_i > 0$.
	Therefore, $E^* \geq P_{min} \cdot \frac{len(\Ja)}{g}$. Thus
	
	\begin{equation}
		\label{eq:lowerEstar}
		 E^{OPT}(\Ja, g) \geq P_{min} \cdot \frac{len(\Ja)}{g}
	\end{equation}
		
	We prove the upper bound of the minimized total energy consumption as following.
	Apply the optimal cost bound in Theorem \ref{def:capacitybounds}, we have $OPT(\Ja, g) \leq len(\Ja)$.
	
	Thus 	
	\begin{equation}
	\label{eq:upper-optimal-energy}
	E^* \leq P_{min} \cdot len(\Ja) +  \sum_{i=1}^{n} e_i .
	\end{equation}
		
	Apply the linear power consumption as in the Equation (\ref{eq:power}) and Equation (\ref{eq:energy}), the energy consumption of each $i$-th virtual machine in period time of [$ts_i, ts_i + d_i$] that denotes as $e_i$ is:
	
	\begin{center}
	$e_i = \int\limits_{ts_i}^{ts_i + d_i} P_j(U_{vm_i})dt = (P_{j}^{max} - P_{j}^{idle}) \cdot U_{vm_i} \cdot d_i$
	\end{center}	
	
	where $U_{vm_i}$ is the percentage of CPU usage of the $i$-th virtual machine on a $j$-th physical machine.
	
	Because any virtual machine always requests CPU usage lesser than or equal to the maximum total capacity CPU of every physical machine, 
	i.e., $U_{vm_i} \leq 1$. 	
	
	\begin{center}
	$\Rightarrow e_i \leq (P_{j}^{max} - P_{j}^{idle}) \cdot d_i $
	\end{center}	
	
	Note that in this proof, all physical machines are identical with same power consumption model thus 
	$P^{max}$ and $P^{idle}$ are the maximum power consumption and the idle power consumption of each physical machine. Thus:
	
	\begin{center}
	$e_i \leq (P^{max} - P^{idle}) \cdot d_i $
	\end{center}
	
	Let $I_i$ is interval of each $i$-th virtual machine, $I_i = [ts_i, ts_i + d_i]$. By the definition the length of interval is $len(I_i) = d_i$ that is duration time of each $i$-th virtual machine. Thus:
	
	\begin{center}
	$e_i \leq (P^{max} - P^{idle}) \cdot len(I_i) $
	\end{center}	
	
	The total energy consumption of $n$ virtual machines is formulated as:
	
	\begin{center}
	$\sum\limits_{i=1}^{n} e_i \leq \sum\limits_{i=1}^{n} [ (P^{max} - P^{idle}) \cdot len(I_i) ]$	
	$\Leftrightarrow \sum\limits_{i=1}^{n} e_i \leq  (P^{max} - P^{idle}) \cdot \sum\limits_{i=1}^{n} len(I_i) $  
	\end{center}
	
	 \begin{equation}
	 \label{eq:sum-energy-n-vms}
	\Leftrightarrow \sum\limits_{i=1}^{n} e_i \leq  (P^{max} - P^{idle}) \cdot len(\Ja).
	 \end{equation}
	
	From Equation (\ref{eq:upper-optimal-energy}), we have:
	
	\begin{center}
	$E^* \leq P_{min} \cdot len(\Ja) +  \sum_{i=1}^{n} e_i $ 	
	$E^* \leq P_{min} \cdot len(\Ja) + (P^{max} - P^{idle}) \cdot len(\Ja)$
	\end{center}
	
	\begin{equation}
		\label{eq:upperEstar1}
		E^* \leq (P_{min} + (P^{max} - P^{idle})) \cdot len(\Ja)
	\end{equation}
	
	By the definition, the unit energy of a physical machine equals to the idle power consumption in the unit time, i.e., $P_{min} = P^{idle}$.
	From the Equation (\ref{eq:upperEstar1}):
	\begin{equation}
		\label{eq:upperEstar2}
		E^* \leq P^{max} \cdot len(\Ja)
	\end{equation}

	\begin{equation}
		\label{eq:upperEstarfinal}
		\Leftrightarrow E^{OPT}(\Ja, g) \leq P^{max} \cdot len(\Ja)
	\end{equation}
	
	From both of two equations (\ref{eq:lowerEstar}) and  (\ref{eq:upperEstarfinal}), we have:
	
	\begin{equation}
		\label{eq:boundedoptimalenergy}
		P_{min} \cdot \frac{len(\Ja)}{g} \leq E^{OPT}(\Ja, g) \leq P^{max} \cdot len(\Ja)
	\end{equation}
	
	We prove the theorem.
	
\end{proof}

\section{Scheduling Algorithms}
\label{sec:algoEMinTRE}
\subsection{EMinTRE-LFT scheduling algorithm}
\label{sec:schedalgo}

\begin{algorithm}[!ht]
\caption{\textbf{: EMinTRE-LFT}: Energy-aware Greedy-based Scheduling Algorithm}\label{alg:eminret}
 \begin{algorithmic}[1]
\Function{EMinTRE-LFT}{}
\State	\textit{Input:} $vmList$ - a list of virtual machines to be scheduled, 
 $hostList$ - a list of physical servers
\State	\textit{Output:} a feasible schedule or $null$
\State	vmList = sortVmListByOrderLastestFinishingTimeFirst( vmList ) \Comment{1}\label{line:sortvmlistbycriteria}
\State  m = hostList.size(); n = vmList.size();
\State T[j] = 0, $\forall j \in [1, m]$ 
\For {$i = 1$ to n} \Comment{on the VMs list}
		\State 	vm = vmList.get(i) 
		\State	allocatedHost = $null$
		\State	T1 = sumTotalHostBusyTime( T ) 
		\State	minRETime = $+\infty$
		\For {$j = 1$ to m}  \Comment{on the hosts list}
         	\State host = hostList.get( $j$ )
         	\State hostVMList = sortVmListByOrder( host.getVms(), order=[finishtime]) 
			\If {host.checkAvailableResource( vm )} \\
				\State preTime = T[ host.id ]
				\State T[ host.id ] = host.estimateHostTotalCompletionTime( vm ) 
				\State T2 = sumTotalHostBusyTime( T )
				\State diffTime = Math.max( T2 - T1, 0)
				\State TRE = EstimateMetricTimeResEff( diffTime, host ) 
				\If {(minTRE $>$ TRE )}
					\State	minRETime = TRE 
					\State	allocatedHost = host 
				\EndIf 
				\State T[ host.id ] = preTime				
				\Comment{ Next iterate over the hostList and choose the host that minimize the value of different time and resource efficiency } 
            \EndIf
		\EndFor 
		
		\If {(allocatedHost != null) }
			\State allocate the $vm$ to the $host$
			\State add the pair of $vm$ (key) and $host$ to the $mapping$
		\Else
			\State "Cannot allocate the virtual machine $vm$."
		\EndIf
\EndFor	
\State	return $mapping$ 
\EndFunction

\State {sumTotalHostBusyTime}({T[]}) = $\sum_{j=1}^{m} T_{j}$ \Comment{T[1...m]: Array of total completion times of $m$ physical servers}

\end{algorithmic}
\end{algorithm}

\begin{algorithm}[ht]
\caption{Estimating the metric for increasing time and resource efficiency}\label{alg:estimretime}
 \begin{algorithmic}[1]
 \Function{EstimateMetricTimeResEff}{}
 \State	\textit{Input:} ($t^{diff}, host$) - $t^{diff}$ is a different time, 
								$host$ is a candidate physical machine
 \State	\textit{Output:} $TRE$ - a value of metric time and resource efficiency
 \State Set $\mathcal{R}$=\{cores, mips, ram, io, netbw, storage\}
 \State $j$ = host.getId(); $n_j$ = host.getVMList();
 \For{$r \in \mathcal{R}$} 
 	\State Calculate the resource utilization, $U_{j,r}$ as in the Equaltion (\ref{eq:resutilization}). 
 \EndFor	
 \State $weights[] \leftarrow$ Read weight of resources from configuration file
 \State Calculate the $TRE_j$ metric of host $j$ as in the equation (\ref{eq:retime})
 \State $D_{j} = \sqrt{ \sum_{r \in \mathcal{R}} ((1 - U_{j,r}) \times w_r)^2 }$
 \State $TRE_j = (\dfrac{t^{diff} \times w_{time}}{T_j^{busy}})^2 + D_j^2 $  \Comment{$w_{time}$ is weight of the different time}
 \State return $TRE_j$
 \EndFunction 
 
\end{algorithmic}
\end{algorithm}

In this section, we present the proposed energy-aware scheduling algorithm, denoted as EMinTRE-LFT, with pseudo-code of EMinTRE-LFT in Algorithm~\ref{alg:eminret}. Algorithm
EMinTRE-LFT has two (2) steps:sorts the list of virtual machines in order decreasing finishing time first. Next, EMinTRE-LFT allocates the first next virtual machine $i$ to 
the first physical machine $M_j$ such that $M_j$ has enough resource to provision the virtual machine $i$ and TRE metric of $M_j$ denoted as $TRE_j$ is minimum. The $TRE_j$ is formulated as in the following equation \ref{eq:retime}. 
The EMinTRE-LFT solves these scheduling problems in time complexity of $\BigO{n \times m \times q}$
where $n$ is the number of VMs to be scheduled, $m$ is the number of physical machines, and $q$ is the maximum number of allocated VMs in the physical machines $M_j, \forall j=1,2,...,m$.

Based on the equation \ref{eq:cpuutilization}, the utilization of a resource $r$ (resource $r$ can be cores, computing power, physical memory, network bandwidth, storage, etc.) of the $M_j$, denoted as $U_{j,r}$, is formulated as:
 \begin{equation}
 \label{eq:resutilization}
 U_{j,r} = \sum\limits_{s \in n_{j}} \dfrac{V_{s,r}}{ H_{j,r}}.
  \end{equation}

{\noindent}where $n_{j}$
is the list of virtual machines that are assigned to the $M_j$, 
 $V_{s,r}$ is the amount of requested resource $r$ of the virtual machine $s$ (note that in our study the value of $V_{s,r}$ is fixed for each user request), 
 and $H_{j,r}$ is the maximum capacity of the resource $r$ in $M_j$.
  
The available resource is presented using diagonal vector, where the $L2$-norm of the diagonal vector (denoted as $D_j$) is formulated as:
\begin{equation}
\label{eq:resefficiency}
D_{j} = \sqrt{( \sum_{r \in \mathcal{R}} ((1 - U_{j,r}) \times w_r)^2 )}
 \end{equation}
{\noindent}where R is the set of resource types in a host ($\mathcal{R}$=\{core, mips, ram, netbw, io, storage\}) and $w_r$ is weight of resource $r$ in a physical machine.

 In this paper, we propose the TRE metric for the increasing total busy time and the $L2$-norm of the diagonal vector ($D_j$) of the physical machine $j$-th that is calculated as:
 \begin{equation}
 \label{eq:retime}
 TRE_j = (\dfrac{t^{diff} \times w_{r=time}}{T_j^{busy}})^2 + D_j^2
  \end{equation}

\section{Performance Evaluation}
\label{sec:experiment}
\subsection{Algorithms}

In this section, we study the following VM allocation algorithms:

\begin{itemize}
\item
PABFD, a power-aware and modified best-fit decreasing heuristic \cite{Beloglazov2012}.
The PABFD sorts the  list of $VM_{i}$ (i=1, 2,..., n) by
their total requested CPU utilization, and 
assigns new VM to any host that has a minimum increase in power consumption.

\item
VBP-Norm-L2, a vector packing heuristics that
is presented as Norm-based Greedy with degree 2 \cite{Panigrahy2011}.
Weights of these Norm-based Greedy heuristics use FFDAvgSum which are $exp(x)$, which is the value of the exponential function at the point $x$, where $x$ is 
average of sum of demand resources (e.g. CPU, memory, storage, network bandwidth, etc.).
VBP-Norm-L2 assigns new VM to any host that has minimum of these norm values.


\item
MinDFT-LDTF: the algorithm sorts list of $VM_{i}$ (i=1, 2,..., n) by their starting time ($ts_{i}$) and respectively by their finished time ($ts_{i}+dur_{i}$), then MinDFT-LDTF allocates each VM (in a given sorted list of VMs) to a host that has a minimum increase in total completion times of hosts as in algorithm MinDFT \cite{HungFDSE2014}. 

\item
EMinTRE-LDTF, the algorithm is proposed in the Section \ref{sec:algoEMinTRE}.

\end{itemize}

\subsection{Methodology}

\begin{table*}[htp]
\caption{Eight (08) VM types in simulations}
\label{tab:vmtype}
\centering
\begin{tabular}{|c|c|c|c|c|c|}
\hline
VM Type                    & \multicolumn{1}{c|}{MIPS} & \multicolumn{1}{c|}{Cores} & \multicolumn{1}{c|}{Memory} & \multicolumn{1}{c|}{Network} & \multicolumn{1}{c|}{Storage} \\ 
                    &  &  & (Unit: MBytes) & (Unit: Mbits/s) & (Unit: GBytes) \\ 
\hline
Type 1 & 2500 & 8 & 6800 & 100 & 1000\\
Type 2 & 2500 & 2 & 1700 & 100 & 422.5\\
Type 3 & 3250 & 8 & 68400 & 100 & 1000\\
Type 4 & 3250 & 4 & 34200 & 100 & 845\\
Type 5 & 3250 & 2 & 17100 & 100 & 422.5\\
Type 6 & 2000 & 4 & 15000 & 100 & 1690\\
Type 7 & 2000 & 2 & 7500 & 100 & 845\\
Type 8 & 1000 & 1 & 1875 & 100 & 211.25\\ \hline
\end{tabular}
\end{table*}

\begin{table*}[htp]
\caption{Information of a typical physical machine (host) 
	with 16 cores CPU (3250 MIPS/core), 136.8 GBytes of available physical memory, 10 Gb/s of network bandwidth, 10 TBytes of storage and idle, maximum power consumption is 175, 250 (W).}
\label{tab:Hostinfo}
\centering
\begin{tabular}{|c|c|c|c|c|c|c|c|}
\hline
Type	& \multicolumn{1}{c|}{MIPS} & \multicolumn{1}{c|}{Cores} & \multicolumn{1}{c|}{Memory} & \multicolumn{1}{c|}{Network} & \multicolumn{1}{c|}{Storage} & \multicolumn{1}{c|}{$P^{idle}$} & \multicolumn{1}{c|}{$P^{max}$}  \\ 
                    &  &  & (Unit: MBytes) & (Unit: Mbits/s) & (Unit: GBytes) & (Unit: Watts) & (Unit: Watts) \\ \hline
M1 & 3250 & 16 & 140084 & 10000 & 10000 & 175 & 250 \\
\hline
\end{tabular}
\end{table*}


\begin{table}
\centering
\caption{The normalized total energy consumption. Simulation results of scheduling algorithms solving scheduling problems with 12681 VMs and 5000 PMs using Feiltelson's parallel workload model \cite{feitelson1996workloadmodel}. Algorithm EMinTRE-LFT wt$X$ has weight of time is equal to $X$ ($X = 1; 0.1; 0.001$).}
	\label{table:simresult-feitelsonmodel1000jobs}
\begin{tabular}{|l|c|c|c|c|}
\hline
Algorithm   & \multicolumn{1}{l|}{Energy} & \multicolumn{1}{l|}{Norm.} & \multicolumn{1}{l|}{Saving Energy}\\ 
	 & (Unit: kWh) & Energy & (+:better;-:worst) \\ \hline
PABFD & 1,055.42 & 1.598 & -60\% \\ 
VBP-Norm-L2 & 1,054.69 & 1.597 & -60\% \\ 
MinDFT-LDTF & 603.90 & 0.915 & 9\% \\ 
Tian-MFFDE & 660.30 & 1.000 & 0\% \\ 
EMinTRE-LFT wt1 & 503.43 & 0.762 & 24\% \\ 
EMinTRE-LFT wt0.01 & 503.43 & 0.762 & 24\% \\ 
EMinTRE-LFT wt0.001 & 503.43 & 0.762 & 24\% \\ \hline
\end{tabular}
\end{table}

\begin{table}
	\centering
	\caption{The normalized total energy consumption. Simulation results of scheduling algorithms solving scheduling problems with 15,201 VMs and 5,000 PMs using Downey97's parallel workload model \cite{downey1998parallel} in the Parallel Workload Archive \cite{feitelsonpwa}. Algorithm EMinTRE-LFT wt$X$ has weight of time is equal to $X$ ($X = 1; 0.1; 0.001$).}
	\label{table:simresult-downey98model1000jobs}
	\begin{tabular}{|l|c|c|c|c|}
		\hline
		Algorithm  & \multicolumn{1}{l|}{Energy} & \multicolumn{1}{l|}{Norm.} & \multicolumn{1}{l|}{Saving Energy}\\ 
				& (Unit: kWh) & Energy & (+:better;-:worst) \\ \hline
		PABFD  & 878.01 & 1.523 & -52.3\% \\
		Norm-VBP-L2  & 876.49 & 1.520 & -52.0\% \\
		Tian-MFFDE  & 576.55 & 1.000 & 0.0\% \\
		MinDFT-LDTF  & 502.61 & 0.872 & 12.8\% \\
		EMinTRE-LFT wt1  & 416.35 & 0.722 & 27.8\% \\
		EMinTRE-LFT wt0.01  & 416.35 & 0.722 & 27.8\% \\
		EMinTRE-LFT wt0.001  & 416.35 & 0.722 & 27.8\% \\ \hline
	\end{tabular}
\end{table}

\begin{table}
	\centering
	\caption{The normalized total energy consumption. Simulation results of scheduling algorithms solving scheduling problems with 8847 VMs and 5000 physical machines (hosts) using Lublin99's parallel workload model \cite{lublin2003workload}}
	\label{table:simresult-lublin99model1000jobs}
	\begin{tabular}{|l|c|c|c|c|}
		\hline
		Algorithm  & \multicolumn{1}{l|}{Energy} & \multicolumn{1}{l|}{Norm.} & \multicolumn{1}{l|}{Saving Energy}\\ 
		& (Unit: kWh) & Energy & (+:better;-:worst) \\ \hline
		PABFD & 460.66 & 1.601 & -60.1\% \\
		Norm-VBP-L2 & 453.23 & 1.575 & -57.5\% \\
		Tian-MFFDE & 287.78 & 1.000 & 0.0\% \\
		MinDFT-LDTF & 263.86 & 0.917 & 8.3\% \\
		EMinTRE-LFT wt0.001 & 232.29 & 0.807 & 19.3\% \\
		EMinTRE-LFT wt0.01 & 232.29 & 0.807 & 19.3\% \\
		EMinTRE-LFT wt1 & 232.29 & 0.807 & 19.3\% \\ \hline		
	\end{tabular}
\end{table}


\begin{figure}[!htp]
	\centering
	\includegraphics[width=0.5\textwidth, height=4cm]{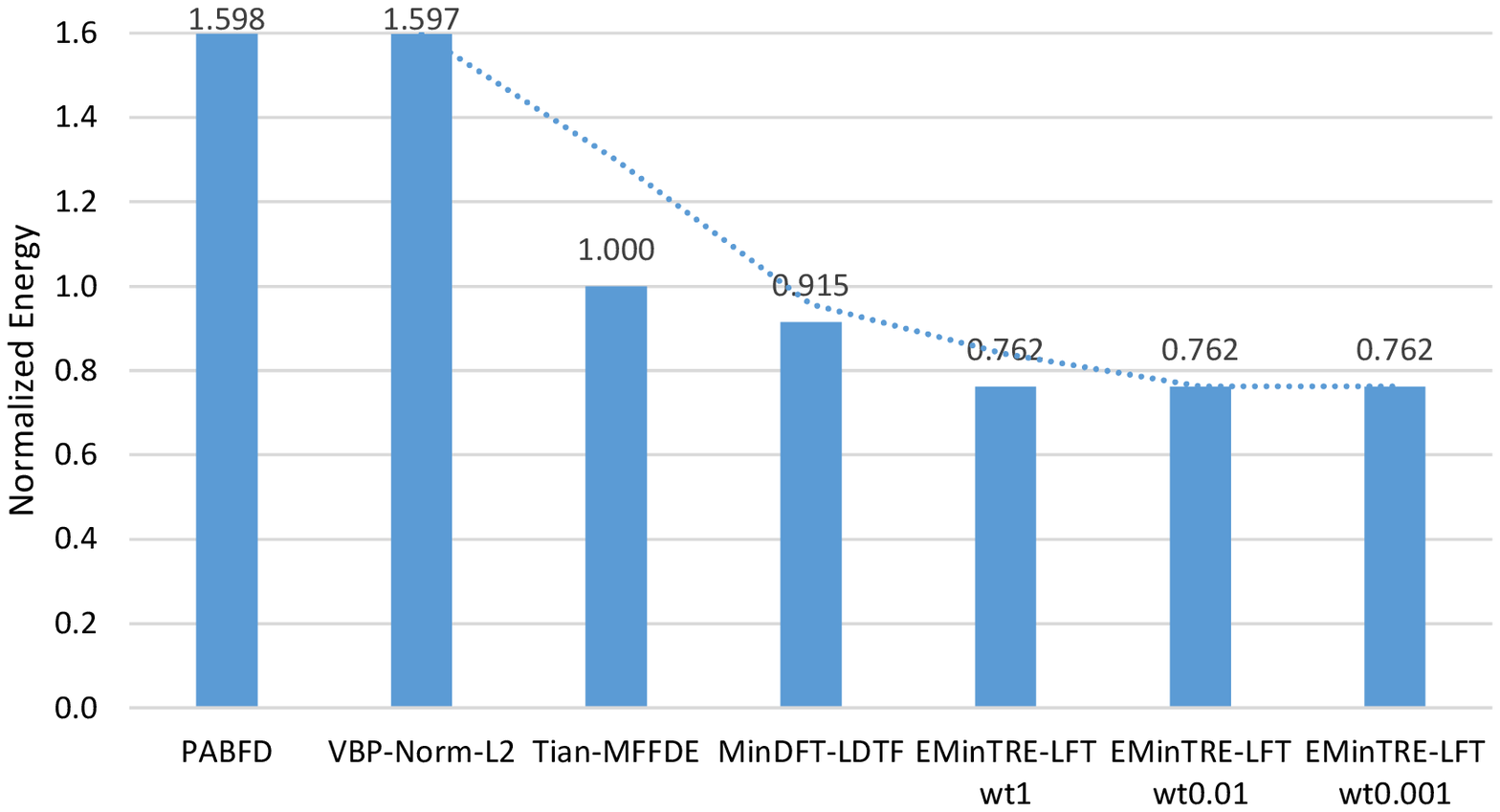}
	\caption{The normalized total energy consumptions compare to Tian-MFFDE. Simulation result for scheduling algorithms with Feitelson's parallel workload model \cite{feitelson1996workloadmodel} in the Parallel Workload Archive \cite{feitelsonpwa} that includes 1,000 jobs have total of 12,681 VMs and 5000 PMs.}
	\label{fig:chart-feitelsonmodel1000jobs}
\end{figure}

\begin{figure}[!htp]
	\centering
	\includegraphics[width=0.5\textwidth, height=4cm]{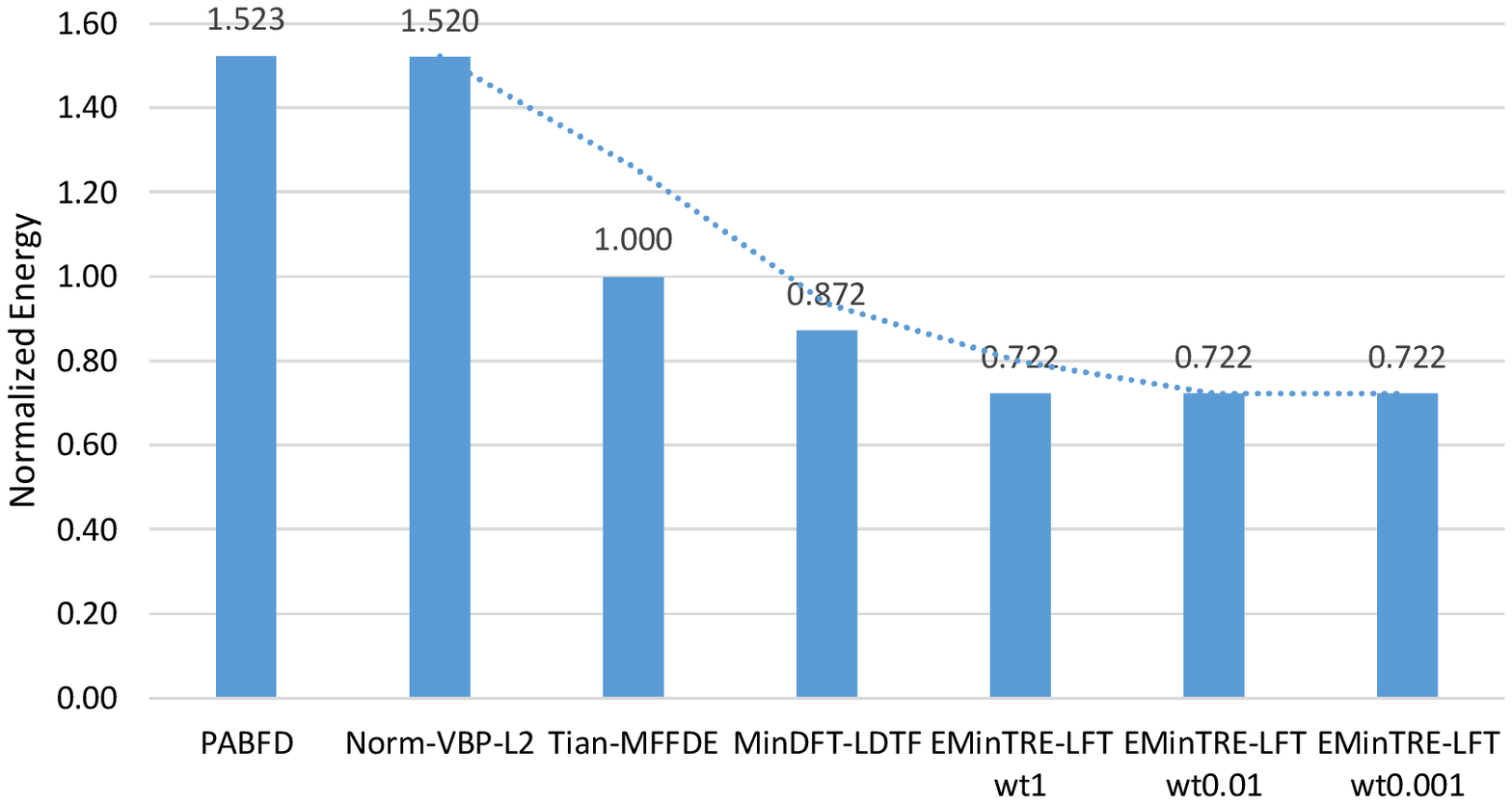}
	\caption{The normalized total energy consumptions compare to Tian-MFFDE. Simulation result for scheduling algorithms with Downey97's parallel workload model \cite{downey1998parallel} in the Parallel Workload Archive \cite{feitelsonpwa} that includes 1,000 jobs have total of 15,201 VMs and 5000 PMs. }
	\label{fig:chart-downey97model1000jobs}
\end{figure}

\begin{figure}[!htp]
	\centering
	\includegraphics[width=0.5\textwidth,height=4cm]{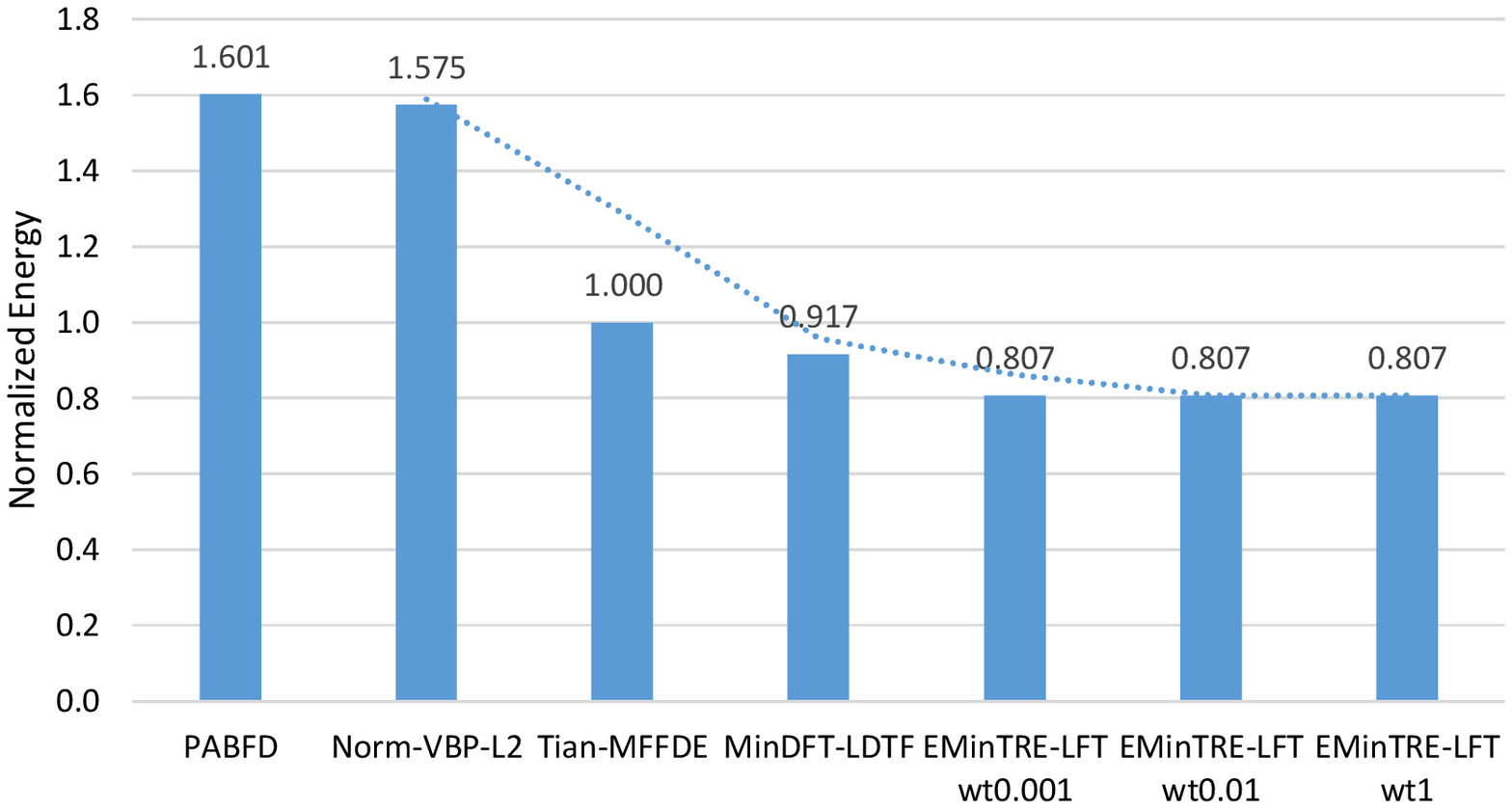}
	\caption{The normalized total energy consumption compare to Tian-MFFDE. Result of simulations with Lublin99's parallel workload model \cite{lublin2003workload} that includes 1,000 jobs have total of 8,847 VMs and 5,000 PMs.}
	\label{fig:chart-energy-lublin99model1000jobs}
\end{figure}

We evaluate these algorithms by simulation using the CloudSim \cite{Cloudsim} to create simulated
 cloud data center systems that have identical physical machines, heterogeneous VMs,
 and with thousands of CloudSim's cloudlets \cite{Cloudsim} (we assume that each
HPC job's task is modeled as a cloudlet that is run on a single VM). 
The information of VMs (and also cloudlets) in these simulated workloads 
is extracted from two parallel job models are Feitelson's parallel workload model \cite{feitelson1996workloadmodel}, 
Downey98's parallel workload model \cite{downey1998parallel} and 
Lublin99's parallel workload model \cite{lublin2003workload}
in Parallel Workloads Archive (PWA) \cite{feitelsonpwa}. When converting from the 
generated log-trace files, each cloudlet's length is a product of the system's 
processing time and CPU rating (we set the CPU rating is equal to included VM's MIPS).
 We convert job's submission time, job's start time (if the start time is missing, then the start time is equal to sum of job's submission time and job's waiting time), 
 job's request run-time, 
 and job's number of processors 
 in job data from the log-trace in PWA \cite{feitelsonpwa}
 to VM's submission time, starting time and duration time, 
 and number of VMs (each VM is created in round-robin in the four types of VMs in Table \ref{tab:vmtype} on the number of VMs). 
 Eight (08) types of VMs as presented in the Table \ref{tab:vmtype} are used in the \cite{WTianMFFDE2013} that are similar to categories in Amazon EC2's 
 VM instances: high-CPU VM, high-memory VM, small VM, and micro VM, etc.. 
 All physical machines are identical and each physical machine is a typical physical machine (Hosts) 
 with 16 cores CPU (3250 MIPS/core), 136.8 GBytes of available physical memory, 10 Gb/s of network bandwidth, 10 TBytes of available storage.
 The minimum and maximum power consumed of each physical machine is 175W and 250W respectively (the minimum power when a PM idle is 175:250 = 70\% of the maximum power consumption as in \cite{Fan2007a}\cite{Beloglazov2012}). 
In the simulations, we use weights as following:
(i) weight of increasing time for mapping a VM to PM: \{0.001, 0.01, 1\};
(ii) weights of computing resources such as number of MIPS per CPU core, physical memory (RAM), network bandwidth, and storage respectively are equally to 1. 
We denoted EMinTRE-LFT wt0.001, EMinTRE-LFT wt0.01 and EMinTRE-LFT wt1 as the total energy consumption of algorithm EMinTRE-LFT in the simulations has weight of increasing time for mapping a VM to PM is \{0.001, 0.01, 1\} respectively.

We choose Modified First-Fit Decreasing Earliest (denoted as Tian-MFFDE) \cite{WTianMFFDE2013} as the baseline because Tian-MFFDE is the best algorithm
in the energy-aware scheduling algorithm to time interval scheduling.
We also compare our proposed VM 
allocation algorithms with 
PABFD \cite{Beloglazov2012} 
because the PABFD is a famous power-aware best-fit decreasing
in the energy-aware scheduling research community, and a vector bin-packing algorithm (VBP-Norm-L2) to show 
the importance of with/without considering VM's starting time and finish time 
in reducing the total energy consumption of VM placement problem.


\subsection{Results and Discussions}
\label{sec:resultsanddiscussions}


The simulation results are shown in the three tables (Table \ref{table:simresult-feitelsonmodel1000jobs}, Table \ref{table:simresult-downey98model1000jobs} and Table \ref{table:simresult-lublin99model1000jobs}) and figures.    
Three (03) figures include Fig. \ref{fig:chart-feitelsonmodel1000jobs}, Fig. \ref{fig:chart-downey97model1000jobs} and Fig. \ref{fig:chart-energy-lublin99model1000jobs} show bar charts comparing energy consumption of VM allocation algorithms that are normalized with the Tian-MFFDE.
None of the scheduling algorithms use VM migration techniques,
and all of them  satisfy the Quality of Service
(e.g. the scheduling algorithm provisions maximum of user VM's requested resources).
We use total energy consumption as the performance metric for
evaluating these VM allocation algorithms.

Using three parallel workload models  \cite{feitelson1996workloadmodel}, \cite{downey1998parallel} and \cite{lublin2003workload} in the Feitelson's Parallel Workloads Archive \cite{feitelsonpwa},
the simulation results show that the proposed EMinTRE-LFT can reduce the total energy consumption of the physical servers by average of 23.7\% 
compared with Tian-MFFDE \cite{WTianMFFDE2013}. 
In addition, EMinTRE-LFT can reduce the total energy consumption of the physical servers by average of 51.5\% and respectively 51.2\% compared with PABFD \cite{Beloglazov2012} and VBP-Norm-L2 \cite{Panigrahy2011}.
Moreover, EMinTRE-LFT has also less total energy consumption than MinDFT-LDTF \cite{HungFDSE2014} in the simulation results. 
\section{Conclusions and Future Work}
\label{sec:concl}
In this paper, we formulated an energy-aware VM allocation problem with multiple resource, fixed interval and non-preemption constraints.
We also discussed our key observation in the VM allocation problem, i.e.,
minimizing total energy consumption is equivalent to
minimize the sum of total completion time of all physical machines (PMs).
Our proposed algorithm EMinTRE-LFT 
can all reduce the total energy consumption
of the physical servers 
compared with the state-of-the-art algorithms in 
simulation results on three parallel workload models of Feitelson's \cite{feitelson1996workloadmodel}, Downey98's \cite{downey1998parallel}, and Lublin99's \cite{lublin2003workload}. 


We are developing the algorithm EMinTRE-LFT into a cloud resource management software (e.g. OpenStack Nova Scheduler). 
 In the future, we would like to evaluate more with the weights of increasing time and $L2$-norm of diagonal vector on available resources.
Additionally, we are working on IaaS cloud systems with heterogeneous physical servers and job requests consisting of multiple VMs using EPOBF \cite{Quang-Hung2014}.
We are studying how to choose the right weights of time and resources 
(e.g. computing power, physical memory, network bandwidth, etc.) in Machine Learning techniques. 


\section*{Acknowledgment}

\bibliographystyle{IEEEtran} 
\bibliography{refs} 

\begin{thebibliography}{10}
\providecommand{\url}[1]{#1}
\csname url@samestyle\endcsname
\providecommand{\newblock}{\relax}
\providecommand{\bibinfo}[2]{#2}
\providecommand{\BIBentrySTDinterwordspacing}{\spaceskip=0pt\relax}
\providecommand{\BIBentryALTinterwordstretchfactor}{4}
\providecommand{\BIBentryALTinterwordspacing}{\spaceskip=\fontdimen2\font plus
\BIBentryALTinterwordstretchfactor\fontdimen3\font minus
  \fontdimen4\font\relax}
\providecommand{\BIBforeignlanguage}[2]{{%
\expandafter\ifx\csname l@#1\endcsname\relax
\typeout{** WARNING: IEEEtran.bst: No hyphenation pattern has been}%
\typeout{** loaded for the language `#1'. Using the pattern for}%
\typeout{** the default language instead.}%
\else
\language=\csname l@#1\endcsname
\fi
#2}}
\providecommand{\BIBdecl}{\relax}
\BIBdecl

\bibitem{Zhang2010a}
\BIBentryALTinterwordspacing
Q.~Zhang, L.~Cheng, and R.~Boutaba, ``{Cloud computing: state-of-the-art and
  research challenges},'' \emph{Journal of Internet Services and Applications},
  vol.~1, no.~1, pp. 7--18, apr 2010.
\BIBentrySTDinterwordspacing

\bibitem{Garg2009}
S.~K. Garg, C.~S. Yeo, A.~Anandasivam, and R.~Buyya, ``{Energy-Efficient
  Scheduling of HPC Applications in Cloud Computing Environments},''
  \emph{CoRR}, vol. abs/0909.1146, 2009.

\bibitem{Le2011}
K.~Le, R.~Bianchini, J.~Zhang, Y.~Jaluria, J.~Meng, and T.~D. Nguyen,
  ``Reducing electricity cost through virtual machine placement in high
  performance computing clouds,'' in \emph{SC}, 2011, p.~22.

\bibitem{Beloglazov2012}
A.~Beloglazov, J.~Abawajy, and R.~Buyya, ``Energy-aware resource allocation
  heuristics for efficient management of data centers for cloud computing,''
  \emph{Future Generation Comp. Syst.}, vol.~28, no.~5, pp. 755--768, 2012.

\bibitem{Fan2007a}
X.~Fan, W.-D. Weber, and L.~Barroso, ``Power provisioning for a warehouse-sized
  computer,'' in \emph{ISCA}, 2007, pp. 13--23.

\bibitem{Quang-Hung2014}
\BIBentryALTinterwordspacing
N.~Quang-Hung, N.~Thoai, and N.~T. Son, ``{EPOBF: Energy Efficient Allocation
  of Virtual Machines in High Performance Computing Cloud},'' \emph{TLDKS XVI},
  vol. LNCS 8960, pp. 71--86, 2014.
\BIBentrySTDinterwordspacing

\bibitem{Tako2012}
I.~Takouna, W.~Dawoud, and C.~Meinel, ``{Energy Efficient Scheduling of
  HPC-jobs on Virtualize Clusters using Host and {VM} Dynamic Configuration},''
  \emph{Operating Systems Review}, vol.~46, no.~2, pp. 19--27, 2012.

\bibitem{Flammini2010}
\BIBentryALTinterwordspacing
M.~Flammini, G.~Monaco, L.~Moscardelli, H.~Shachnai, M.~Shalom, T.~Tamir, and
  S.~Zaks, ``{Minimizing total busy time in parallel scheduling with
  application to optical networks},'' \emph{Theoretical Computer Science}, vol.
  411, no. 40-42, pp. 3553--3562, Sep. 2010.
\BIBentrySTDinterwordspacing

\bibitem{WTianMFFDE2013}
\BIBentryALTinterwordspacing
W.~Tian and C.~S. Yeo, ``{Minimizing total busy time in offline parallel
  scheduling with application to energy efficiency in cloud computing},''
  \emph{Concurrency and Computation: Practice and Experience}, vol.~27, no.~9,
  pp. 2470--2488, jun 2013.
\BIBentrySTDinterwordspacing

\bibitem{Panigrahy2011}
R.~Panigrahy, K.~Talwar, L.~Uyeda, and U.~Wieder, ``{Heuristics for Vector Bin
  Packing},'' Microsoft Research, Tech. Rep., 2011.

\bibitem{HungFDSE2014}
N.~Quang-Hung, D.-K. Le, N.~Thoai, and N.~T. Son, ``{Heuristics for
  Energy-Aware VM Allocation in HPC Clouds},'' \emph{Future Data and Security
  Engineering (FDSE 2014)}, vol. 8860, pp. 248--261, 2014.

\bibitem{feitelson1996workloadmodel}
D.~G. Feitelson, ``Packing schemes for gang scheduling,'' in \emph{Job
  Scheduling Strategies for Parallel Processing}.\hskip 1em plus 0.5em minus
  0.4em\relax Springer, 1996, pp. 89--110.

\bibitem{downey1998parallel}
A.~B. Downey, ``A parallel workload model and its implications for processor
  allocation,'' \emph{Cluster Computing}, vol.~1, no.~1, pp. 133--145, 1998.

\bibitem{lublin2003workload}
U.~Lublin and D.~G. Feitelson, ``The workload on parallel supercomputers:
  modeling the characteristics of rigid jobs,'' \emph{Journal of Parallel and
  Distributed Computing}, vol.~63, no.~11, pp. 1105--1122, 2003.

\bibitem{feitelsonpwa}
D.~G. Feitelson, ``{Parallel Workloads Archive},'' (retrieved on 31 Januray
  2014), http://www.cs.huji.ac.il/labs/parallel/workload/.

\bibitem{kovalyov2007fixed}
M.~Y. Kovalyov, C.~Ng, and T.~E. Cheng, ``Fixed interval scheduling: Models,
  applications, computational complexity and algorithms,'' \emph{European
  Journal of Operational Research}, vol. 178, no.~2, pp. 331--342, 2007.

\bibitem{Angelelli20113650}
\BIBentryALTinterwordspacing
E.~Angelelli and C.~Filippi, ``On the complexity of interval scheduling with a
  resource constraint,'' \emph{Theoretical Computer Science}, vol. 412, no.~29,
  pp. 3650--3657, 2011.
\BIBentrySTDinterwordspacing

\bibitem{Knauth2012}
\BIBentryALTinterwordspacing
T.~Knauth and C.~Fetzer, ``{Energy-aware scheduling for infrastructure
  clouds},'' in \emph{4th IEEE International Conference on Cloud Computing
  Technology and Science Proceedings}.\hskip 1em plus 0.5em minus 0.4em\relax
  IEEE, Dec. 2012, pp. 58--65.
\BIBentrySTDinterwordspacing

\bibitem{Chen2014}
\BIBentryALTinterwordspacing
L.~Chen and H.~Shen, ``{Consolidating complementary VMs with
  spatial/temporal-awareness in cloud datacenters},'' in \emph{IEEE INFOCOM
  2014 - IEEE Conference on Computer Communications}.\hskip 1em plus 0.5em
  minus 0.4em\relax IEEE, Apr. 2014, pp. 1033--1041.
\BIBentrySTDinterwordspacing

\bibitem{BelBuLZA2010Taxonomy}
A.~Beloglazov, R.~Buyya, Y.~C. Lee, and A.~Zomaya, ``{A Taxonomy and Survey of
  Energy-Efficient Data Centers and Cloud Computing Systems},'' \emph{Advances
  in Computers}, vol.~82, pp. 1--51, 2011.

\bibitem{orgerie2014survey}
\BIBentryALTinterwordspacing
A.-C. Orgerie, M.~D. de~Assuncao, and L.~Lefevre, ``{A survey on techniques for
  improving the energy efficiency of large-scale distributed systems},''
  \emph{ACM Computing Surveys}, vol.~46, no.~4, pp. 1--31, Mar. 2014.
\BIBentrySTDinterwordspacing

\bibitem{Hameed2014}
\BIBentryALTinterwordspacing
A.~Hameed, A.~Khoshkbarforoushha, R.~Ranjan, P.~P. Jayaraman, J.~Kolodziej,
  P.~Balaji, S.~Zeadally, Q.~M. Malluhi, N.~Tziritas, A.~Vishnu, S.~U. Khan,
  and A.~Zomaya, ``{A survey and taxonomy on energy efficient resource
  allocation techniques for cloud computing systems},'' pp. 1--24, Jun. 2014.
\BIBentrySTDinterwordspacing

\bibitem{IvonaSurvey2014}
\BIBentryALTinterwordspacing
T.~Mastelic, A.~Oleksiak, H.~Claussen, I.~Brandic, J.-M. Pierson, and A.~V.
  Vasilakos, ``Cloud computing: Survey on energy efficiency,'' \emph{ACM
  Comput. Surv.}, vol.~47, no.~2, pp. 33:1--33:36, Dec. 2014.
\BIBentrySTDinterwordspacing

\bibitem{Cloudsim}
R.~N. Calheiros, R.~Ranjan, A.~Beloglazov, C.~A.~F. De~Rose, and R.~Buyya,
  ``Cloudsim: a toolkit for modeling and simulation of cloud computing
  environments and evaluation of resource provisioning algorithms,''
  \emph{Softw., Pract. Exper.}, vol.~41, no.~1, pp. 23--50, 2011.

\end{thebibliography}

\end{document}